\documentclass[aps,prl,floats,twocolumn]{revtex4}

\usepackage{epsfig}

\newcommand{\be}{\begin{equation}}
\newcommand{\ee}{\end{equation}}
\newcommand{\bea}{\begin{eqnarray}}
\newcommand{\eea}{\end{eqnarray}}
\newcommand{\bt}{\begin{tabular}}
\newcommand{\et}{\end{tabular}}
\newcommand{\ba}{\begin{array}}
\newcommand{\ea}{\end{array}}

\begin{document}




\title{Mpemba effect and phase transitions in the adiabatic cooling
of water before freezing}

\author{S. Esposito\footnote{Corresponding author, Salvatore.Esposito@na.infn.it}}
\author{R. De Risi} \author{L. Somma}
\affiliation{Dipartimento di Scienze Fisiche, Universit\`{a} di
Napoli
``Federico II'' and \\ Istituto Nazionale di Fisica Nucleare, Sezione di Napoli \\
Complesso Universitario di Monte S. Angelo, Via Cinthia, I-80126 Napoli, Italy}


\begin{abstract}
An accurate experimental investigation on the Mpemba effect (that
is, the freezing of initially hot water before cold one) is
carried out, showing that in the adiabatic cooling of water a
relevant role is played by supercooling as well as by phase
transitions taking place at $6 \pm 1^{\rm o}$C, $3.5 \pm 0.5^{\rm
o}$C and $1.3 \pm 0.6^{\rm o}$C, respectively. The last
transition, occurring with a non negligible probability of 0.21,
has not been detected earlier. Supported by the experimental
results achieved, a thorough theoretical analysis of supercooling
and such phase transitions, which are interpreted in terms of
different ordering of clusters of molecules in water, is given.
\end{abstract}

\maketitle




\noindent A well-known phenomenon such as that of the freezing of
water has attracted much interest in recent times due to some
counter-intuitive experimental results \cite{mpemba} and the
apparent lacking of a generally accepted physical interpretation
of them \cite{kell}, \cite{woj}, \cite{supercooling}, \cite{katz}.
These results consist in the fact that, {\it many times},
initially hot water freezes more quickly than initially cold one,
a phenomenon which is now referred to as the Mpemba effect (for a
short historical and scientific survey see the references in
\cite{supercooling}). The observations sound counter-intuitive
when adopting the naive, simple view according to which initially
hot water has first to cool down to the temperature of the
initially cold one, and then closely follow the cooling curve of
the last one. The effect takes place even for not pure water, with
solutions or different liquids (the original Mpemba observation
occurred when he tried to make an ice cream).

Several possible physical phenomena, aimed to explain such
observations, have been proposed in the literature, mainly
pointing out that some change in water should occur when heated
\cite{kell} \cite{katz}.

However, such explanations cannot be applied if some precautions
are taken during the experiments (whilst the Mpemba effect has
been observed even in these cases) and, in any case, calculations
do not seem to support quantitatively the appearance of the effect
(see references in \cite{supercooling}).

Some novel light has been introduced in the discussion, in our opinion, in Ref.
\cite{supercooling}, where the Mpemba effect has been related to the occurrence of
supercooling both in preheated and in non-preheated water. Initially hot water seems to
supercool to a higher local temperature than cold water, thus spontaneously freezing
earlier. As a consequence, such a scenario, apparently supported by experimental
investigations, points toward a statistical explanation of the effect, neither the time
elapsed nor the effective freezing temperature being predictable.

Here, we prefer to  face the problem by starting from what is
known about the {\it freezing} process, rather than the {\it
cooling} one.

In general it is known that, for given values of the thermodynamic
quantities (for example the volume and the energy), a physical
system may exist in a state in which it is not homogeneous, but it
breaks into two or more homogeneous parts in mutual equilibrium
between them. This happens when stability conditions are not
fulfilled, so that a phase transition occurs; it is, for example,
just the case of water that, at the pressure $p$ of 1 atm and at
temperature $T$ of $0^{\rm o}$C, becomes unstable.

When liquid water is cooled, the average velocities of its
molecules decreases but, even if the temperature goes down to
$0^{\rm o}$C (the fixed temperature where liquid and solid phases
coexist) or lower, this is not a sufficient condition for freezing
to start. In fact, in order that ice begins to form, first of all
some molecules of the liquid water should arrange in a
well-defined order to form a minimum crystal and this, in the
liquid state, may happen only randomly. Second, such starting {\it
nucleus} has to attract further molecules in the characteristic
locations of the crystalline structure of ice, by means of the
interaction forces of the nucleus with the non-ordered molecules
in the liquid. Nucleation and crystal growth processes are both
favored at temperatures lower than $0^{\rm o}$C, so that
supercooling of liquid water is generally required before its
effective freezing. In fact, in pure water, only molecules in the
liquid with statistically lower velocities can arrange the initial
nucleus and, furthermore, only slow moving molecules are able to
join that cluster and put their kinetic energies into potential
energy of bond formation. When ice begins to form, these molecules
are removed from those attaining to the given Maxwell distribution
for the liquid water, so that the average speed becomes larger,
and the temperature of the system rises to $0^{\rm o}$C
(obviously, the temperature is set at the value where the
continuing exchange of molecules is equal in terms of those
joining and those leaving the formed crystal surface).

Thus supercooling is, {\it de facto}, a key ingredient in the
freezing process, although supercooled water exists in a state of
precarious equilibrium (water is in a metastable state). Minor
perturbations such as impurities  or other can trigger the sudden
appearance of the stable crystalline phase for the whole liquid
mass, again with the release of the entire crystallization heat
(melting heat) which increases the temperature of the freezing
liquid to the normal $0^{\rm o}$C one.

In general, when a system is in a metastable state, sooner or
later it will pass to another stable state. In water, density and
entropy fluctuations favor the formation of crystallization nuclei
but, if the liquid constitutes a stable state, such nuclei are
always unstable and will disappear with time being. However, since
the fluctuations become more pronounced the lower the temperature,
if water is supercooled, for sufficiently large nuclei they will
result to be stable and grew with time, becoming freezing centers.
The starting of the phase transition is thus {\it determined} by
the probability of appearance of those nuclei, and the reported
Mpemba effect could be simply a manifestation of this process.

We have calculated just this probability $P$ as function of the
absolute temperature $T$ of the metastable phase (the one at which
the nucleus is in equilibrium with the liquid), obtaining the
following result \footnote{We do not give the details of such
calculations; the interested reader may follow those reported in
section 162 of Ref. \cite{landau} for a similar case.}:
\begin{equation}
P = \frac{\alpha}{T_\ast} \, {\rm exp} \left\{ - \beta \,
\frac{T_\ast^2}{(T-T_\ast)^2} \right\} . \label{1}
\end{equation}
Here $T_\ast$ is the equilibrium temperature of the liquid-solid
phase, $\alpha$ is a dimensionless normalization factor and
$\beta$ is a constant whose expression is given by
\begin{equation}
\beta = \frac{16 \pi \tau^3 v^2}{3 Q^2 k T_\ast} , \label{2}
\end{equation}
where $\tau$ is the surface tension, $v$ the molecular volume of
the crystallization nucleus, $Q$ the molecular heat of the
transition from the metastable phase to the nucleus phase, and $k$
is the Boltzmann constant. Just to give an idea of the macroscopic
value of the constant $\beta$, let us note that $\tau^3 v^2 =
W_{\rm surf}^3$ is the cube of the work done by the surface
forces, and by assuming that $Q \sim k T_\ast$ we may write:
\begin{equation}
\beta \simeq \frac{16 \pi}{3} \left( \frac{W_{\rm surf}}{Q}
\right)^3 , \label{3}
\end{equation}
that is the constant $\beta$ is ruled by the ratio $W_{\rm surf}/Q$.

The probability $P$ has a minimum at the liquid-solid equilibrium
temperature $T_\ast$ and increases for decreasing temperature, as
expected. From the formulae above it is clear that the probability
for nucleation, and thus the onset of the freezing process as
well, is enhanced if the work done by the surface forces (or the
surface tension itself) is lowered in some way. In normal daily
conditions when a commercial refrigerator is employed, this is
easily induced in two simple ways: either by the presence of
impurities, when solutions (such as an ice cream solution, as in
the Mpemba case) are used as the freezing liquid instead of pure
water, or by fluctuations of the external pressure or temperature,
caused in the commercial refrigerator itself. This explains why no
appreciable supercooling is observed in normal situations.
Obviously, the most direct way to induce freezing in supercooled
water is to introduce an external body in it, in order to directly
lower the surface tension.

We have thus performed an accurate experimental investigation,
accounting for a total of about one hundred runs, aimed to clarify
the phenomenology of the Mpemba effect and its interpretation. In
the first part of our experiments we have tested all the above
qualitative predictions about supercooling, by studying the
cooling and freezing of tens of cm$^3$ of normal water in a
commercial refrigerator, in daily operation conditions. The key
point, in fact, is not to obtain the most favorable physical
conditions, employing sophisticated setups, but rather to
reproduce the Mpemba conditions, that is adiabatic cooling (with
commercial refrigerators) of not extremely small quantities of
water. We have used an Onofri refrigerator for the cooling of
double distilled water and a NiCr-Ni thermocouple as a temperature
sensor (Leybold 666193), interfaced with a Cassy Lab software for
data acquisition.

For fixed temperatures of the cryostat we have indeed observed
supercooling in our samples, with the freezing occurring just
along the lines predicted above. In particular, during the
supercooling phase we have induced a number of small perturbations
in our samples, namely, variations of external pressure or
temperature, mechanical perturbations or introduction of an
external macroscopic body (a glass thermometer held at the same
temperature of the sample). In {\it all} these cases we have
registered the sudden interruption of the supercooling phase and a
practically instantaneous increase of the temperature to the value
of $0^{\rm o}$C, denoting the starting of the freezing process.
Conversely, if no perturbation is induced (or takes places) the
water reached an equilibrium with the cryostat at temperatures up
to about $-30^{\rm o}$C (lasting also for several thousands of
seconds).

We have then verified that when the freezing process started from
the supercooling phase, the Mpemba effect took place with a
probability in agreement with that reported in Ref.
\cite{supercooling}.

\begin{table}
\footnotesize
\begin{tabular}{|c|c|c|c|c|}
\hline %
& $V$= 20 cm$^3$ & $V$= 50 cm$^3$ & $V$= 65 cm$^3$ & $V$= 80
cm$^3$ \\ \hline %
$P_{SC}$ & $1$ & $0.28$ & $0$ & $0.46$ \\
\hline \hline
 & $T_c= -8^{\rm o}$C & $T_c= -14^{\rm o}$C & $T_c= -22^{\rm o}$C
 & $T_c= -26^{\rm o}$C \\ \hline
$P_{SC}$ & $0.75$ & $0.50$ & $0$ & $0.11$ \\ \hline
\end{tabular}
\caption{Probabilities for the occurrence of supercooling for
different volumes $V$ of the sample and different temperatures of
the cryostat $T_c$.} \label{t1}
\end{table}

In about half (with a total probability of 0.47) of the runs
performed we have detected a supercooling phase. In Table \ref{t1}
we report the observed probability $P_{SC}$ for the occurrence of
supercooling for different volumes $V$ of the water sample and for
different temperatures $T_c$ of the cryostat. We find the data to
be fitted by a straight line, denoting (in the range considered) a
linear decreasing of $P_{SC}$ for decreasing temperatures of the
cryostat and for increasing volumes of the samples, this
probability reaching the maximum $P_{SC}=1$ for $T_c=0^{\rm o}$C
(and $V=0$).

An interesting feature of what we have observed is the sensible
appearance of iced water in our samples. In fact, when
supercooling did not occur, the ice started to form around the
walls of the beaker, while the inner parts were still in a liquid
form, as usually expected. Instead the immediate freezing of
supercooled water involved the {\it whole} sample, this showing a
very peculiar symmetric form. We have used cylindrical beakers
with the temperature sensor in their periphery, near the walls;
the observed structure was a pure radial (planar) one, with no
liquid water and radial filaments of ice from the center of the
beakers to the walls (in one case we have been also able to take a
low resolution picture of this, before its destruction outside the
refrigerator).

\begin{figure}
\begin{tabular}{cc}
\mbox{\epsfysize=4truecm %
\epsffile{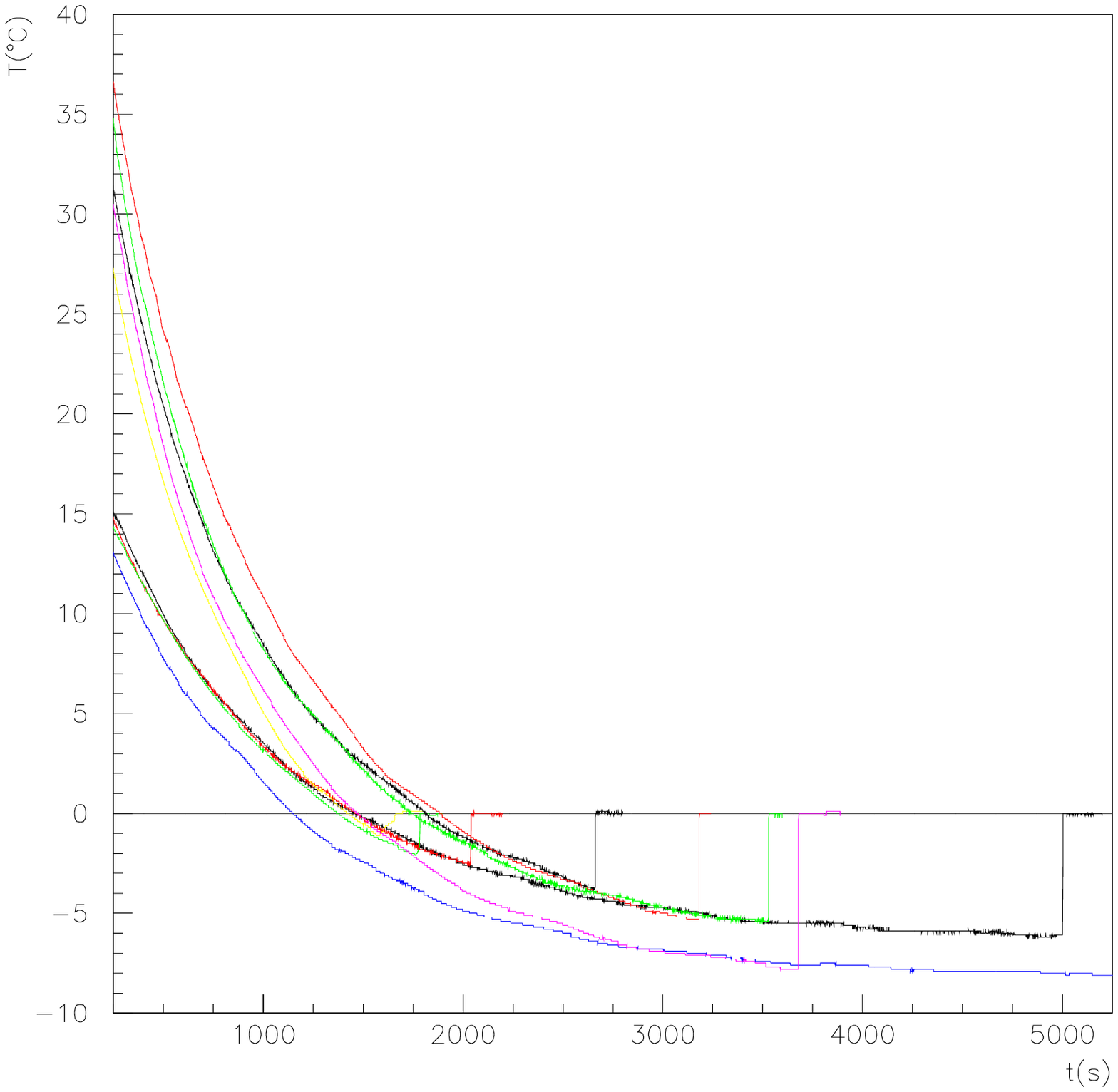}} %
&
\epsfysize=4truecm %
\epsffile{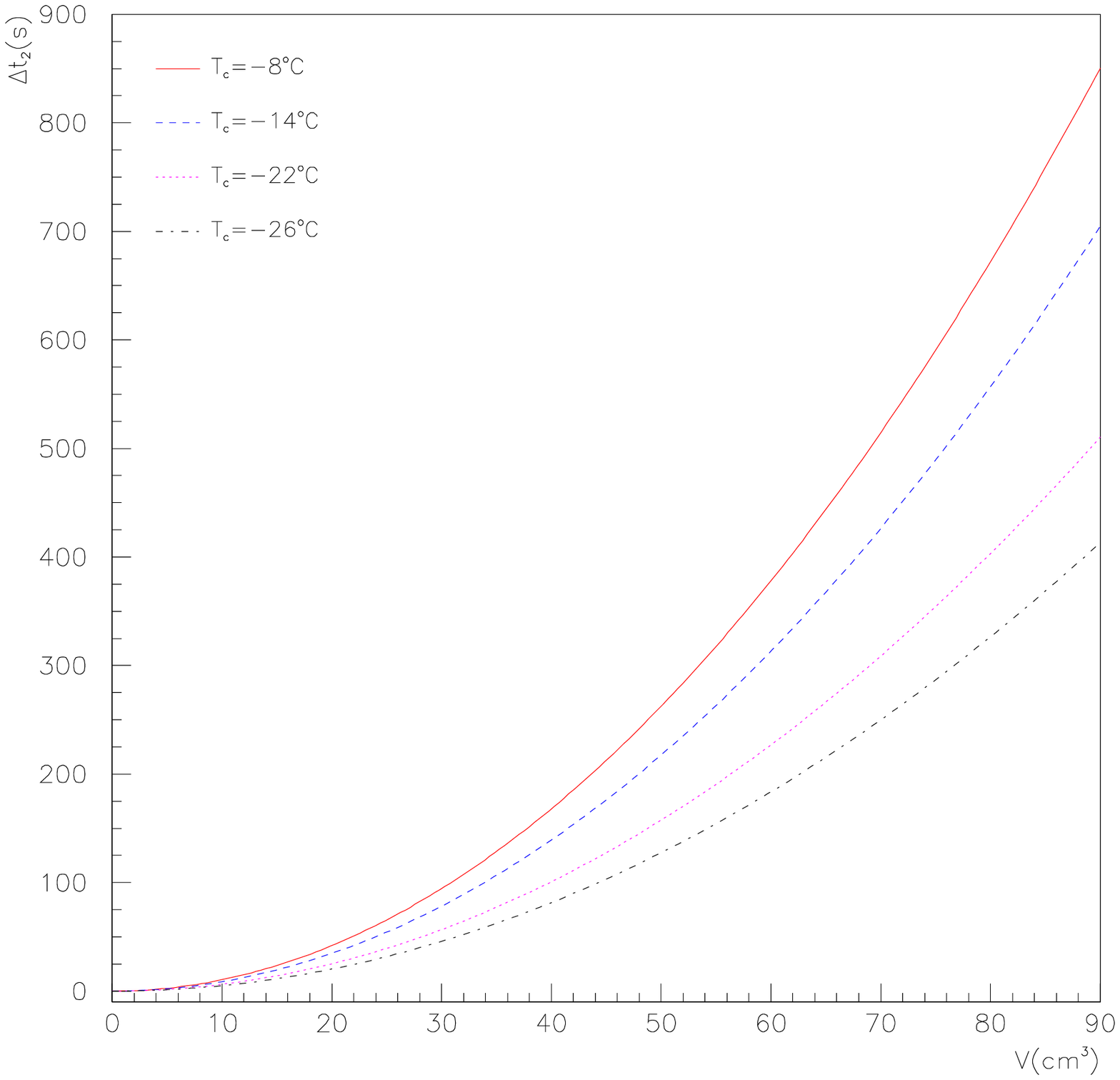} %
\end{tabular}
\caption{Left: Cooling curves for $V=20$ cm$^3$ and $T_c=-8^{\rm
o}$C. \\ Right: The fitted time duration of the phase transition
at $3.5^{\rm o}$C as function of the volume $V$ of the samples,
for different temperatures $T_c$ of the cryostat.}
\label{f1} \label{f2}
\end{figure}

However, although supercooling plays a relevant role in the
manifestation of the Mpemba effect, the things are made more
complicated by the occurrence of other statistical effects before
the temperature of the water reaches the value of $0^{\rm o}$C.
This comes out when an accurate measurement of the cooling curves
is performed (some examples of what we have obtained during the
second part of our experiments are reported in Fig. \ref{f1}).

According to a simple naive model, the heat exchange from the
water sample (at initial temperature $T_0$) to the cryostat (at
fixed temperature $T_c$) is described by the equation
\begin{equation}
C \, {\rm d} T \; = \; \delta \left( T_c - T_0 \right) {\rm d} t ,
\label{ph1}
\end{equation}
where $C$ and $\delta$ are the thermal capacity and the heat
conductivity of the water, respectively. Thus by solving the
differential equation in (\ref{ph1}), the following expression for
the temperature as function of time $t$ is obtained:
\begin{equation}
T \; = \; T_c - \left( T_c - T_0 \right) {\rm e}^{-t/\tau} ,
\label{ph2}
\end{equation}
where $\tau = C/\delta$ is a time constant measuring the cooling
rate of the sample. However, although the overall dependence of
$T$ on time is that expressed by Eq. (\ref{ph2}), our experimental
data clearly reveal the presence of three transition points before
freezing (or supercooling), where $\tau$ changes its value. This
transitions occur at temperatures $T_1=6 \pm 1^{\rm o}$C, $T_2=3.5
\pm 0.5^{\rm o}$C and $T_3=1.3 \pm 0.6^{\rm o}$C with a
probability of $P_1=0.11$, $P_2=0.84$ and $P_3=0.21$,
respectively. The time duration $\Delta t$ of each phase
transition, during which the temperature keeps practically
constant \footnote{In some cases we have been able to observe also
a van der Waals-like profile of $T(t)$ at the transition point
(metastable state), instead of only the mean constant value of
$T$.}, depends on the volume of the sample and on the temperature
of the cryostat. The data we have collected are summarized in
Table \ref{t2}. For the phase transition at $T_2$ these data show
a linear increase of $\Delta t_2$ with $T_c$ and a quadratic one
with $V$; in Fig. \ref{f2} we give the fitting curves
corresponding to best fit function $\Delta t_2 = (a+b T_c)V^2$.
Instead, for the other two phase transitions no sufficient data
are available in order to draw  any definite conclusion on the
dependence on $V$ and $T_c$, though $\Delta t_1$ and $\Delta t_3$
appear to be shorter than $\Delta t_2$.

\begin{table}
\footnotesize
\begin{tabular}{|c|c|c|c|c|}
\hline %
\multicolumn{5}{|c|}{$T_c=-8 \pm 2 ^{\rm o}$C} \\ \hline
& $V$= 20 cm$^3$ & $V$= 50 cm$^3$ & $V$= 65 cm$^3$ & $V$= 80 cm$^3$ \\
\hline
$\Delta t_1$ (s) & $7 \pm 1$ & & &  \\
$\Delta t_2$ (s) & $11 \pm 6$ & $220 \pm 100$ & $500 \pm 170$
 & $630 \pm 160$ \\
$\Delta t_3$ (s) & $12 \pm 6$ & & $70 \pm 30$ & \\ \hline
\end{tabular}
\begin{tabular}{|c|c|c|c|c|}
\hline %
\multicolumn{5}{|c|}{$T_c=-14 \pm 2 ^{\rm o}$C} \\ \hline
& $V$= 20 cm$^3$ & $V$= 50 cm$^3$ & $V$= 65 cm$^3$ & $V$= 80 cm$^3$ \\
\hline
$\Delta t_1$ (s) & & & & $37 \pm 1$ \\
$\Delta t_2$ (s) & $8 \pm 3$ & $130 \pm 80$ & $480 \pm 160$
 & $500 \pm 60$ \\
$\Delta t_3$ (s) & $7 \pm 4$ & & & \\ \hline
\end{tabular}
\begin{tabular}{|c|c|c|c|c|}
\hline %
\multicolumn{5}{|c|}{$T_c=-22 \pm 1 ^{\rm o}$C} \\ \hline
& $V$= 20 cm$^3$ & $V$= 50 cm$^3$ & $V$= 65 cm$^3$ & $V$= 80 cm$^3$ \\
\hline
$\Delta t_1$ (s) & & $63 \pm 1$ & & $7 \pm 1$ \\
$\Delta t_2$ (s) & & $170 \pm 100$ & & $130 \pm 70$ \\
$\Delta t_3$ (s) & & & & \\ \hline
\end{tabular}
\begin{tabular}{|c|c|c|c|c|}
\hline %
\multicolumn{5}{|c|}{$T_c=-26 \pm 1 ^{\rm o}$C} \\ \hline
& $V$= 20 cm$^3$ & $V$= 50 cm$^3$ & $V$= 65 cm$^3$ & $V$= 80 cm$^3$ \\
\hline
$\Delta t_1$ (s) & & $3.5 \pm 0.7$ & &  \\
$\Delta t_2$ (s) & & $3 \pm 1$ & $320 \pm 70$ & $210 \pm 170$ \\
$\Delta t_3$ (s) & & $200 \pm 70$ & $1 \pm 1$ & \\ \hline
\end{tabular}
\caption{Time duration of the phase transitions at $6 ^{\rm o}$C
($\Delta t_1$), $3.5 ^{\rm o}$C ($\Delta t_2$) and $1.3 ^{\rm o}$C
($\Delta t_3$) for different volumes $V$ of the sample and
different temperatures of the cryostat $T_c$.}%
\label{t2}
\end{table}


The occurrence of these phase transitions is likely related to the
formation of more or less ordered structures in water, resulting
from the competition between long-range density ordering and local
bond ordering maximizing the number of local bonds \cite{tanaka}.
The anomalous density maximum at about $4^{\rm o}$C (which we
observe here at $T_2=3.5 \pm 0.5^{\rm o}$C) is, for example,
explained just in term of this: as water is cooled, the local
specific volume increases due to the progressive increase in
tetrahedral order, so that the entropy, that always decreases upon
cooling, at $4^{\rm o}$C becomes anticorrelated with the volume,
resulting in an inversion (from positive to negative) of the
thermal expansion coefficient and a corresponding density maximum
\cite{deben}. Similar explanations in terms of different ordering
could apply also to the other two transitions we have observed,
but an exhaustive discussion of them, which would require more
experimental data, is beyond the scope of this Letter. We only
note that, while the first transition at $T_1=6\pm 1^{\rm o}$C
seems related to the effect observed in Ref. \cite{katera} at
$8^{\rm o}$C, to the best of our knowledge no other author has
reported the one at $T_3=1.3 \pm 0.6^{\rm o}$C (which, as
mentioned, occurs with an appreciable probability of 0.21).

\begin{table}
\footnotesize
\begin{tabular}{|c|c|c|c|c|}
\hline %
 & $T_c= -8^{\rm o}$C & $T_c= -14^{\rm o}$C & $T_c= -22^{\rm o}$C
 & $T_c= -26^{\rm o}$C \\ \hline
$\tau_1$ (s) & $600 \pm 110$ & $680 \pm 100$ & $1000 \pm 110$ & $950 \pm 190$ \\
$\tau_2$ (s) & $1080 \pm 260$ & $1060 \pm 170$ & $530 \pm 90$ & $570 \pm 3$ \\
$\tau_3$ (s) & $1590 \pm 930$ & $1520 \pm 730$ & $270 \pm 50$ & $220 \pm 80$ \\
$\tau_4$ (s) & $620 \pm 480$ & $500 \pm 180$ & $150 \pm 30$ & $640
\pm 490$  \\ \hline
\end{tabular}
\caption{Time constants $\tau_1$ ($T<T_1$), $\tau_2$
($T_1<T<T_2$), $\tau_3$ ($T_2<T<T_3$), $\tau_4$ ($T>T_3$) of the
cooling curves before and after the three phase transitions
detected, for different temperatures of the cryostat $T_c$.}
\label{t3}
\end{table}

The observed mean values of the four time constants of the cooling
curves, before and after the three phase transitions, are reported
in Table \ref{t3} for different values of $T_c$. All the time
constants are approximately {\it independent} on the volume $V,$
in disagreement with the naive model discussed above which
predicts an increase of $\tau$ with the thermal capacity. Instead
they depend linearly on $T_c$, showing a negative slope for
$\tau_1$ and positive ones for $\tau_2, \tau_3, \tau_4$ and a
finite value for $T_c=0^{\rm o}$C. Note that (in the naive model)
the ratios of the different time constants, at fixed volumes, give
the (inverse) ratios of the heat conductivities in the different
ordered phases (all these ratios decrease with the cryostat
temperature), which are directly related to microscopic quantities
like the size and average velocity of the ordered clusters of
molecules in water.

Coming back to the Mpemba effect, it is easy to see that Eq.
(\ref{ph2}) predicts that, for constant $\tau$, initially hot
water reaches the freezing point {\it later} than initially cold
water. However, from what just discussed, in general this could be
no longer true if the time constant changes its value during the
cooling process (the slope of the cooling curves changes), or
phase transitions before freezing occur (with time durations
sufficiently long/short). In addition to these effects, the
reaching of the freezing point does not automatically guarantees
the effective starting of the freezing process, since relevant
supercooling may take place, thus statistically causing the
freezing of initially hot water {\it before} cold one.

From the data we have collected we have verified that, for given
$V$ and $T_c$, in many cases no inversion between the cooling
curves happens before the freezing point, irrespective of the
change in the value of $\tau$ or the time duration of the phase
transitions. Nevertheless we have as well realized that this is
mainly due to the not very large difference between the initial
temperatures of the samples, and in few cases (among those studied
by ourselves) it cannot be applied, the largest effect causing the
inversion being the phase transition at $T_2$.

In conclusion our experimental results, and their interpretation
reported here, clearly point out the statistical nature of the
Mpemba effect (as already realized in \cite{supercooling}), whose
explanation is given in terms of transitions between differently
ordered phases in water and supercooling. The very detection of
such phenomena seems to require the cooling to be adiabatic (as
fulfilled in our experiment, as well as in those performed by
other authors \cite{supercooling}), since for non adiabatic
processes (for example, in fused salt) the coexistence of local
solid nuclei in the liquid phase has been observed \footnote{We
are indebted with M. Villa for having pointed out this to us.}.

An unexpected novel transition at $T_3=1.3\pm 0.6^{\rm o}$C has
been as well detected with a non negligible probability, calling
for further accurate investigation in order to achieve a more
complete understanding of the unique properties of water.



{\it Acknowledgements:} Interesting discussions with G. Salesi and
M. Villa are kindly acknowledged.


\end{document}